\documentclass[fleqn,12pt]{wlscirep}
\usepackage[utf8]{inputenc}
\usepackage[T1]{fontenc}
\usepackage{bm}
\usepackage{graphicx}

\usepackage{bm}

\def\eg{e.g.,~}
\def\ie{i.e.,~}

\usepackage[nameinlink]{cleveref} 
\usepackage{url}
\usepackage{color}
\usepackage{xcolor}
\hypersetup{
 colorlinks  = true, 
 urlcolor   = black, 
 linkcolor  = blue, 
 citecolor  = blue 
}

\newcommand{\figref}[1]{Fig.~\ref{fig:#1}}

\usepackage{ulem}
\usepackage{bm}

\definecolor{purple}{RGB}{170, 0, 255}

\let\emph\relax 
\DeclareTextFontCommand{\emph}{\it}

\author[1]{Sadamori Kojaku}
\author[2,3]{Laurent H\'ebert-Dufresne}
\author[4]{Enys Mones}
\author[4,5]{Sune Lehmann}
\author[1,6,*]{Yong-Yeol Ahn}
\affil[1]{Center for Complex Networks and Systems Research, Luddy School of Informatics, Computing and Engineering, Indiana University, Bloomington, IN 47408, USA}
\affil[2]{Vermont Complex Systems Center, University of Vermont, Burlington, VT 05405, USA}
\affil[3]{Department of Computer Science, University of Vermont, Burlington, VT 05405, USA}
\affil[4]{DTU Compute, Technical University of Denmark, 2800 Kgs. Lyngby, Denmark}
\affil[5]{Center for Social Data Science, University of Copenhagen, 1353 Copenhagen K, Denmark}
\affil[6]{Indiana University Network Science Institute, Indiana University, Bloomington, IN 47408, USA}
\affil[*]{email: yyahn@iu.edu}

\date{\today}

\title{The effectiveness of backward contact tracing in networks} 

\begin{abstract} 

Discovering and isolating infected individuals is a cornerstone of epidemic control~\cite{cohen2003efficient,barthelemy2005dynamical,christakis2010social,park2020coronavirus,patient31wp,Gilbert2020,Mattioli2020}.
Because many infectious diseases spread through close contacts, contact tracing is a key tool for case discovery and control~\cite{eames2003contact,lloyd2005superspreading,klinkenberg2006effectiveness,andre2007transmission,glasser2011modeling,peak2017comparing,ferretti2020quantifying,Aleta2020}. 
However, although contact tracing has been performed widely, the mathematical understanding of contact tracing has not been fully established and it has not been clearly understood what determines the efficacy of contact tracing.
Here, we reveal that, compared with ``forward'' tracing---tracing {\it to} whom disease spreads, ``backward'' tracing---tracing {\it from} whom disease spreads---is profoundly more effective. 
The effectiveness of backward tracing is due to simple but overlooked biases arising from the heterogeneity in contacts. 
Using simulations on both synthetic and high-resolution empirical contact datasets, we show that even at a small probability of detecting infected individuals, strategically executed contact tracing can prevent a significant fraction of further transmissions. 
We also show that---in terms of the number of prevented transmissions per isolation---case isolation combined with a small amount of contact tracing is more efficient than case isolation alone.
By demonstrating that backward contact tracing is highly effective at discovering super-spreading events, we argue that the potential effectiveness of contact tracing has been underestimated.
Therefore, there is a critical need for revisiting current contact tracing strategies so that they leverage all forms of biases. 
Our results also have important consequences for digital contact tracing because it will be crucial to incorporate the capability for backward and deep tracing while adhering to the privacy-preserving requirements of these new platforms. 
\end{abstract}

\begin{document}

\flushbottom
\maketitle

\thispagestyle{empty}

\section{Introduction}\label{sec:introduction} 

Mass quarantine has shown its effectiveness in controlling the epidemic outbreak during the COVID-19 pandemic, but with a considerable social and economic cost~\cite{Gilbert2020,Mattioli2020}.
Once the initial outbreak has been suppressed, it is critical to manage resurgence in order to avoid uncontrolled spreading and another lockdown.
Infection does not occur spontaneously but does so through close physical contacts.
Therefore, contact tracing---tracing and isolating close contacts of infected individuals to prevent further transmission---is a potent intervention measure for successful epidemic control~\cite{eames2003contact,lloyd2005superspreading,klinkenberg2006effectiveness,andre2007transmission,christakis2010social,glasser2011modeling,peak2017comparing,ferretti2020quantifying,Aleta2020}.
For instance, contact tracing has played a critical role in ending the SARS outbreak in 2003 and discovered many super-spreading events in the COVID-19 pandemic~\cite{glasser2011modeling,park2020coronavirus}.
However, because traditional contact tracing is labor-intensive and slow, its efficacy and cost-benefit trade-offs have been questioned~\cite{armbruster2007contact,hellewell2020feasibility}.
Therefore, \textit{digital contact tracing} that leverages mobile devices may allow more swift and efficient contact tracing, potentially overcoming the limitations of the traditional contact tracing~\cite{ferretti2020quantifying}. 

Regardless of whether it is performed in person or digitally, contact tracing in practice often discovers super-spreading events, which are abundant in many epidemics~\cite{lloyd2005superspreading}.
A famous example from the COVID-19 pandemic would be the `Shincheonji Church' associated with the `Patient $31$' in South Korea~\cite{patient31wp}. 
The patient was the first positive case from the church-event, which was later identified---via contact tracing---to be the single biggest super-spreading event in South Korea. This single super-spreading event eventually caused more than $5,000$ cases, accounting for \textit{more than half} of South Korea's total cases during that time~\cite{patient31wp}. 
As illustrated in this case, super-spreading events are the norm rather than the exception~\cite{lloyd2005superspreading}, and these events are often discovered through contact tracing efforts~\cite{andre2007transmission, park2020coronavirus}.

The contact tracing's ability to detect super-spreading events can be, in part, attributed to the ``friendship paradox''~\cite{feld1991your}.
The friendship paradox states that your friends tend to have more friends than you, because the more friends someone has, the more often they show up in someone's friend list. 
Now, because a disease is transmitted through contact ties, the disease preferentially reaches individuals with many contacts who can potentially cause super-spreading events. Beyond being an interesting piece of trivia, this insight has proven useful for epidemic surveillance and control~\cite{christakis2010social}. 
Individuals with many social contacts such as celebrities and politicians are in many ways ideal sentinel-nodes for epidemic outbreaks~\cite{cohen2003efficient, christakis2010social,barthelemy2005dynamical, lloyd2005superspreading}.

Here we argue that contact tracing is assisted by an additional statistical bias in social networks. 
This bias is leveraged when the contact tracing is executed {\it backward} to identify the source of infection (parent). 
This is because the more offsprings (infections) a parent has produced, the more frequently the parent shows up as a contact. 
Both biases can be at play at the same time, and thus their effects are additive, resulting in an exceptional efficacy of backward contact tracing at identifying super-spreaders and super-spreading events. 

A leading factor that determines the strengths of these statistical biases is the structural properties of the underlying contact network itself, in particular, the heterogeneity of the degree (\ie the number of contacts). 
Heterogeneous networks, where the number of contacts varies significantly among individuals, have a larger variance in the degree, which in turn produces a stronger friendship paradox effect. 
Real networks are known to be heterogeneous \cite{Albert2002,Dorogovtsev2003,Pastor-Satorras2004}, with strong implications for epidemiology because these properties alter the fundamental nature of the epidemic dynamics in the form of, for instance, vanishing epidemic threshold~\cite{pastor2001epidemic}, hierarchical spreading~\cite{barthelemy2004velocity}, and large variance in individual's reproductive number~\cite{lloyd2005superspreading} as well as the final outbreak size~\cite{hebert2020beyond}. 

Here, we analyze the statistical biases that backward contact tracing leverages. 
Using simulations on both synthetic and empirical contact network data, we show that strategically executed contact tracing can be highly effective and efficient at controlling epidemics. 
Our results call not only for the incorporation of contact tracing as a more crucial part of the epidemic control strategy, but crucially for the implementation of backward-facing contact tracing protocols both in traditional and digital contact tracing programs to fully leverage the biases afforded by empirical network structures.

\section{Results}\label{sec:results} 

\subsection{Bias owing to the friendship paradox}
\begin{figure}
  \centering
  \includegraphics[width=\columnwidth]{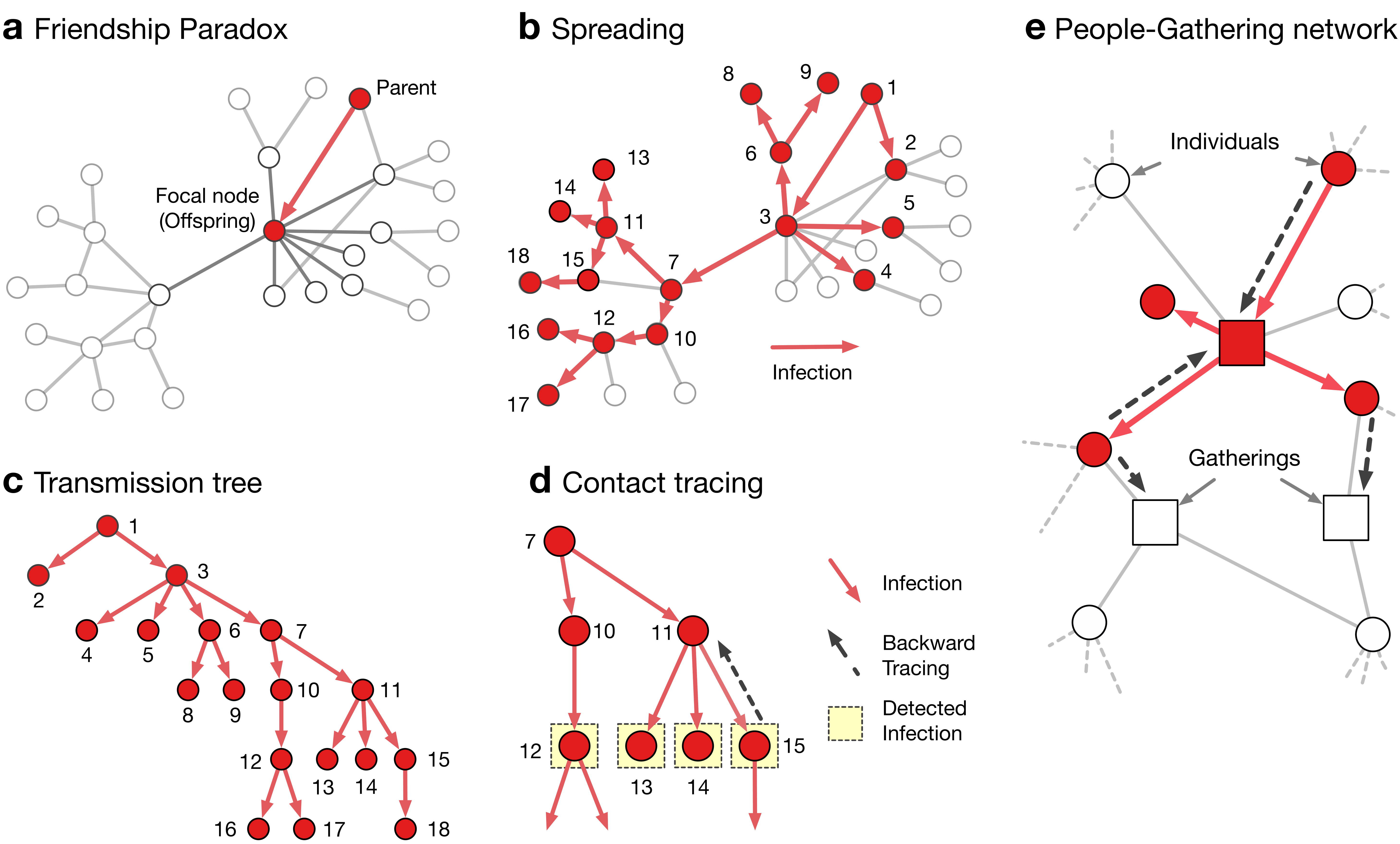} 
  \includegraphics[width=\columnwidth]{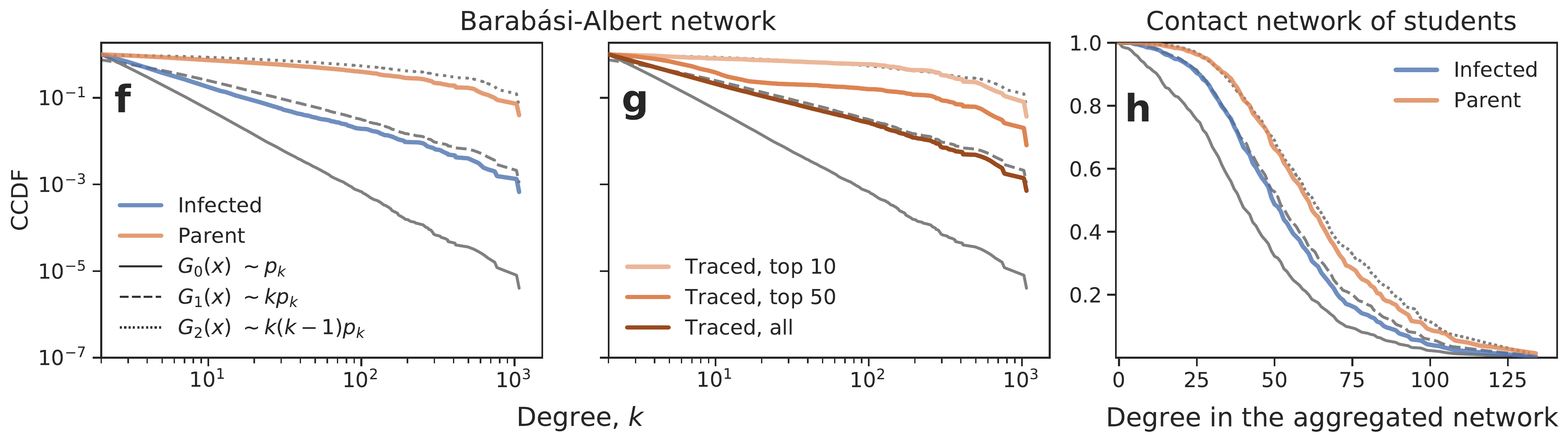}
  \caption{{\bf a}. Schematic illustration of a contact network. A transmission event occurs from a `parent' to a `focal node' (or an offspring). 
  {\bf b}. The disease spreads from an infected node to its neighbors through edges in networks.
{\bf c}. The spread of disease can be represented as a transmission tree with directed edges from parents to offsprings.
{\bf d}. Backward tracing is likely to sample parents with many offsprings, \eg node 11 is more likely to be sampled than node 10 by backward tracing.
  {\bf e}. Contact tracing can be also conducted for a bipartite network of people and gatherings. 
  As to the contact network, a high-degree gathering is more likely to be ``infected'' and to be traced with the same logic.  
  {\bf f}. As a proof of concept, we simulate the SIR model on the Barab\'asi-Albert network composed of $250,000$ nodes. We sample the infected nodes with probability $0.1$ and trace their parents at time $t=0.5$.
  The blue and orange lines indicate the degree distributions for the sampled nodes and their parents, which follow $G_1$ and $G_2$, respectively.
  {\bf g}. The frequency-based contact tracing---isolating the most frequently traced nodes---can reach nodes with a degree similar to the parents without knowing who-infects-whom.
  {\bf h}. The bias owing to backward tracing is present even in a relatively homogeneous network.
  We simulate the SEIR model on a temporal contact network of university students and sample all infected nodes and their parents. 
  The infected and parent nodes have the degree distributions that closely follow $G_1$ and $G_2$ for the unweighted aggregated network, respectively.   
  }%
  \label{fig:diagrams}
\end{figure}

Face-to-face contacts between people can be represented as a network, where a node is a person and an edge indicates a contact between two persons.
When a node in the network is infectious, the disease can be transmitted to the neighbors through its edges (\figref{diagrams}a).
A node with many edges is likely to be one of the neighbors and thus has a high chance of infection.
This is the friendship paradox described above~\cite{feld1991your}. 
In other words, ``you'' are a random node having $k$ contacts drawn from a distribution $p_k$, whereas ``your friends'' are those having $k'$ contacts drawn proportionally to $k'p_{k'}$. 
The friendship paradox aggravates epidemic outbreaks because individuals with many contacts are preferentially infected, and spread the infection to many \cite{pastor2001epidemic, barthelemy2004velocity, newman2005threshold}.

Formally, if we sample a node at random, the distribution of degree (\ie the number of contacts) is given by $\{p_k\}$, which can be expressed as the probability generating function (PGF), \ie 
\begin{equation}
G_0(x) = \sum_k p_k x^k.
\end{equation}
The PGF is a polynomial representation of the degree distribution; for example, the average degree can be calculated using a derivative $\langle k \rangle = \sum_k kp_k = G'_0(1)$.
Now, consider that a node is infected and the disease is transmitted through an edge chosen at random.
Then, the disease is $k$ times more likely to reach a node with degree $k$ than a node with degree $1$. 
Therefore, the number of other contacts (i.e, \textit{excess} degree; $k-1$) found at the end of that contact is generated by
\begin{equation}
G_1(x) = \frac{1}{\langle k \rangle}\sum_k kp_k x^{k-1},
\end{equation}
where $\langle k \rangle$ is a normalization constant. Note that the average excess degree is larger than or equal to the average degree, $G'_1(1) \geq G'_0(1)$ (Friendship paradox). 

This property can be leveraged by the so-called ``acquaintance sampling'' strategy, where one randomly samples individuals and then samples their ``friends'' by following contacts~\cite{cohen2003efficient, christakis2010social}.
Because the acquaintance sampling can preferentially sample hubs in a network even without knowing its whole structure, it has been shown to help early detection of an outbreak as well as efficient control of the disease~\cite{cohen2003efficient, christakis2010social}.

\subsection{Bias owing to backward tracing}

An often overlooked fact about contact tracing is that there are two \textit{directions} that the contact tracing can lead to discovery of infected individuals. 
The first is following the direction of the transmission---\textit{to whom} the transmission may have occurred---and the other is reaching to the parent---\textit{from whom} the transmission occurred. 
The difference has a profound implication on the statistical nature of the sampling. 

Disease spreading can be represented as a tree composed of edges from parents to offsprings (\figref{diagrams}c). 
If we follow the transmission edge to the offsprings of a node, we are sampling with the bias owing to the friendship paradox ($\sim kp_k$). 
However, when we trace back to the parent, another statistical bias comes into play. 
Imagine someone who has spread the disease to $k$ individuals (\eg node 11 in \figref{diagrams}d) and another infected individual who only spreads the disease to one individual (node 10). 
If we sample infected individuals (one of nodes 12--15) and follow a transmission edge \textit{back} to the parent, we are likely to reach the one who has more offspring (node 11).
Formally, if we trace back to the parent, the number of the other offsprings for the parent is generated by
\begin{equation}
G_2(x) = \frac{G'_1(x)}{G'_1(1)} = \frac{1}{\sum_k k(k-1)p_k} \sum_k k (k-1) p_k x^{k-2}.
\end{equation}
The contact tracing samples a parent having $k-2$ degree (\ie the number of other offsprings) with a probability proportional to $k(k-1)p_k$ ($\sim k^2p_k$)---a bias stronger than acquaintance sampling ($\sim kp_k$). 
To illustrate this in practice, we simulate the Susceptible-Infectious-Recovered (SIR) model on a degree heterogeneous network generated by the Barab\'asi-Albert (BA) model \cite{barabasi1999emergence} (See Methods on the parameters of the SIR model).
At an early stage ($t=0.5$), the degree distribution for all infected nodes and that for parents closely follow the distributions proportional to ${kp_k}$ and ${k(k-1) p_k}$, respectively (\figref{diagrams}f).

Backward tracing needs information about the direction from which the infection occurs.
However, except for a few diseases~\cite{meyers2006400}, the direction of transmission is not clear in practice.
Still, we can preferentially sample super-spreading parents (events) by leveraging the bias owing to backward tracing.
Because a super-spreader or super-spreading event infects many individuals,
they would appear as a common contact or visited location of many infected individuals.
For example, in \figref{diagrams}d, node 11 is a common neighbor for 3 infected nodes and hence would appear 3 times more frequently than node 10.
The bias can be leveraged by the {\it frequency-based} contact tracing, where we trace and isolate the most frequent nodes in the contact list.
For the BA network, the frequency-based contact tracing samples nodes with a degree similar to the parents without knowing the direction of transmissions (\figref{diagrams}g).

\subsection{Effectiveness of contact tracing for heterogeneous networks}

The backward tracing leverages the two sampling biases attributed to the heterogeneity in the degree distributions.
Therefore, we hypothesize that contact tracing is highly effective in degree heterogeneous networks.
As a proof of concept, we simulate epidemic spreading using the SIR model on a network with a power-law degree distribution. 
The network is generated by the BA model composed of $250,000$ nodes with minimum degree 2~\cite{barabasi1999emergence} (see Simulating epidemic spreading in Methods for parameter values). 
Although the SIR model simulated on the BA networks, in many respects, differ from epidemic spreading in empirical social networks~\cite{stumpf2012,broido2019,hellewell2020feasibility}, it demonstrates that contact tracing can leverage the sampling biases arising from the heterogeneity.

We intervene epidemic spreading from $t=0.5$ by detecting and isolating newly infected individual at the time of infection with probability $p_s$ (\ie probability of detecting infection).
Then, from each detected individual, we add each contact (\ie neighbor) to a contact list with probability $p_t$ (\ie probability of successful tracing).
At every interval of $\Delta t = 0.25$, we isolate the most frequent $n$ nodes in the contact list and then clear the list. Note that contact tracing with $p_t=0$ is equivalent to case isolation, \ie we discover and isolate newly infected nodes with probability $p_s$ but do not trace close contacts.
We model the contact tracing as preventing infections to all nodes rooted from the isolated nodes in the transmission tree. 

\begin{figure}
  \centering
  \includegraphics[width=\linewidth]{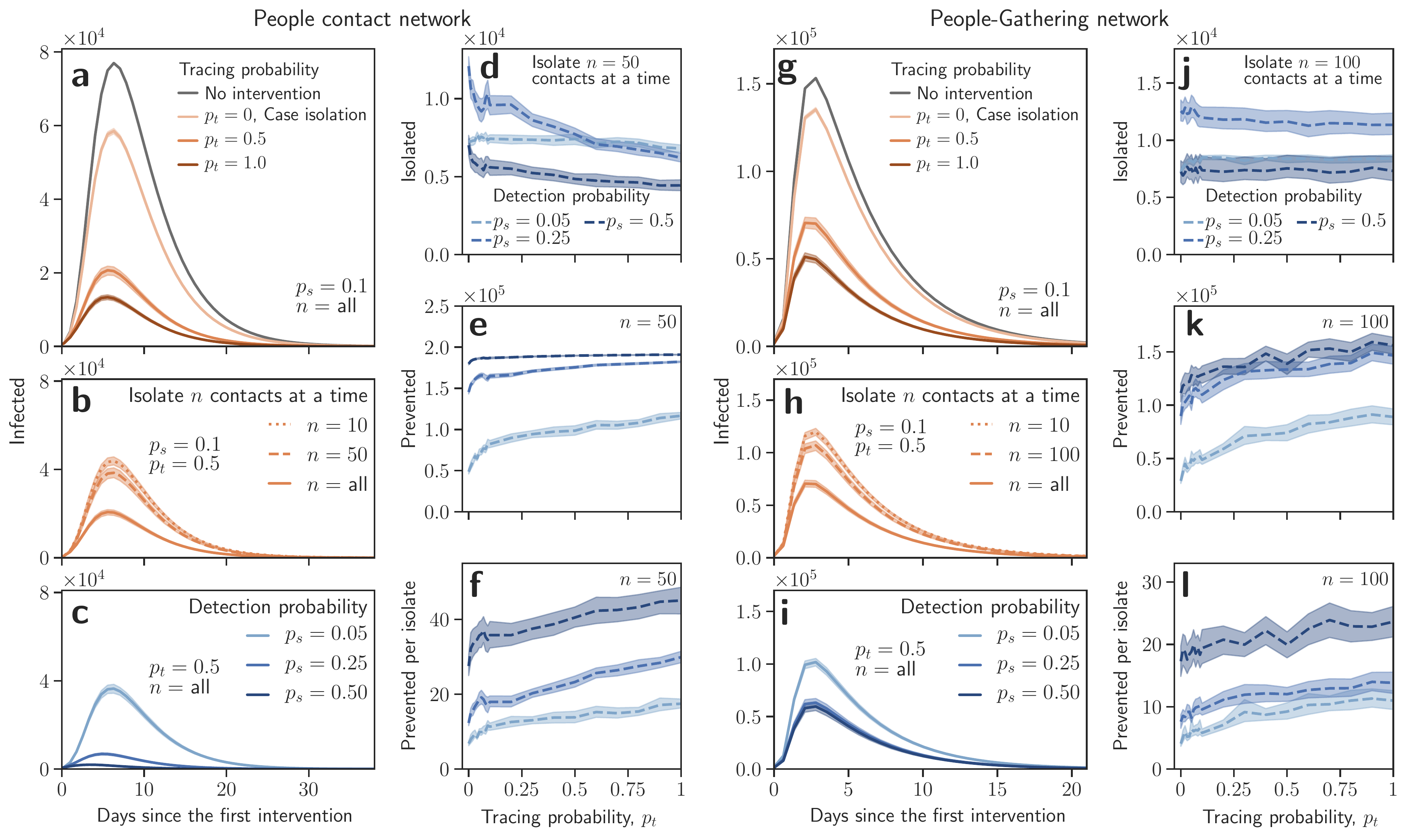}
  \caption{
    Effectiveness of contact tracing for networks with a heterogeneous degree distribution. 
    ({\bf a}--{\bf f}) People contact networks. 
    ({\bf g })--({\bf l}). People-gathering networks. 
    The people networks and people-gathering networks are generated by the BA and the configuration model, respectively.
    {\bf a}. Contact tracing lowers the peak of infection by more than 50\% of that for case infection. 
    {\bf b}. The effectiveness stands out even if we cannot trace all but few nodes. 
    {\bf c}. The efficacy of contact tracing is substantially enhanced when the detection probability is increased.  
    {\bf d--f}. Compared to case isolation ($p_t = 0$), contact tracing ($p_t>0$) isolates fewer nodes while preventing more cases. 
    Therefore, contact tracing is highly cost-efficient in terms of the number of prevented cases per isolation.
    {\bf g}--{\bf i}. Contact tracing is also highly effective for people-gathering networks.
    {\bf j}--{\bf l}. Compared to case isolation, contact tracing isolates a comparable number of cases while preventing more cases, leading to a higher cost-efficiency.
    Each point indicates the average value for $30$ simulations.
    The translucent band indicates the 95\% confidence interval estimated by a bootstrapping with $10^4$ resamples.
  }%
  \label{fig:synthe_effectiveness}
\end{figure}

The disease infects roughly 30\% of nodes at the peak of infection (\figref{synthe_effectiveness}a). 
The peak can be reduced by more than 70\% with contact tracing for $p_t\geq 0.5$ (\figref{synthe_effectiveness}a).  
Even a small amount of extra isolations through contact tracing (\eg $n=10$ from the population of $250,000$) is still effective in flattening the curve of infections (\figref{synthe_effectiveness}b).
The effectiveness is more pronounced when we can identify more infected nodes, \eg by increasing the number of testing (\figref{synthe_effectiveness}c).

Contact tracing isolates {\it fewer} nodes in total while preventing more cases than case isolation, resulting in a high cost-efficiency in terms of the number of prevented cases per isolation (\figref{synthe_effectiveness}d--f).
This might appear to be counter-intuitive because contact tracing isolates extra nodes (\ie contacts) in addition to case isolation.
However, because this additional isolation by contact tracing preferentially targets those who are at high risk, they, in turn, prevents many subsequent transmission events, reducing the total number of isolation. 


Outbreak investigation can be considered as contact tracing for `gatherings' (e.g.~the closure of churches, grocery markets, or any spontaneous gatherings; see \figref{diagrams}e)~\cite{sekara2016fundamental}.
Note that the privacy-preserving contact tracing protocols such as DP-3T~\cite{dp3t} can be used to detect spreading events that happened in gatherings and notify risk information for those who joined the gatherings. Moreover, the people-gathering structure is found in high temporal resolution proximity data \cite{sekara2016fundamental} and is stable because human mobility often follows regular routines \cite{sekara2016fundamental, song2010limits, bagrow2012mesoscopic}. 

Contact tracing is effective at detecting the gatherings with super-spreading events for the same reason as for super-spreaders; gatherings with $k$ participants are detected with a probability roughly proportional to $k^2$ (see People--gathering networks in Methods).
To test its effectiveness, we generate synthetic people-gathering networks composed of $200,000$ person-nodes and $50,000$ gathering-nodes with a power-law distribution of exponent $\beta = -3$ using the configuration model \cite{Fosdick2018}. Then, we run the SIR simulations on the network (see Simulating epidemic spreading in Methods). Contact tracing is executed from $t\geq 0.1$ in the same way as to people contact network. 

As in the case of people contact networks, contact tracing substantially reduces the peak of infections (\figref{synthe_effectiveness}g).
The effectiveness stands out even if we do not isolate all but only $10$ gatherings from a population of $200,000$ people and $50,000$ gatherings (\figref{synthe_effectiveness}h and i).
Contact tracing isolates a comparable number of nodes as case isolation while preventing more infections, yielding a higher cost-efficiency (\figref{synthe_effectiveness}j--l).

\subsection{Contact tracing on temporal contact network of students}

A virus can easily spread in a densely connected population where people routinely have face-to-face contact with each other such as students participating in the same class~\cite{gemmetto2014,darbon2019}, and workers in dorms~\cite{sadaranguni2017}.
Without physical distancing, epidemic control is extremely difficult. If large gatherings (\eg classes) are prohibited, there may not be strong heterogeneity in terms of the offspring distribution (no super-spreading events). In such a case, would contact tracing be useful at all?

We test the effectiveness of contact tracing for a temporal contact network of $567$ university students, which is constructed using the physical contact data collected in the Copenhagen Network Study \cite{sapiezynsky2019interaction}.
The physical contacts are estimated by smartphones at 5 minutes resolutions.
This network, as it only captures the infections among a specific population and neglects others, has a fairly homogeneous degree distribution, with the maximum degree 42 at five minutes resolutions. 

The epidemic spreading is simulated using a more empirically-grounded model---Susceptible-Exposed-Infectious-Recovered (SEIR) model---which reflects the fact that many infectious diseases have an incubation period before being infectious~\cite{hellewell2020feasibility} (see Methods for data prepossessing and parameters for the SEIR model). 
Even in this fairly homogeneous network, the sampling biases are present; for instance, the parents of infected nodes have a larger degree than the infected nodes in the aggregated network (see \figref{diagrams}h).

We carry out contact tracing on the third day and onward in the same way as for the synthetic networks except how we compile the contact list.
We detect newly infected individuals with probability $p_s$ at the time of infectious. 
Then, with a probability $p_t$, a close contact for each detected individual is traced and added to the contact list; we consider a node as a close contact if and only if it has contact with the detected individual for at least one hour in the previous seven days. The contact tracing is carried out at every interval of $24$ hours.

\begin{figure}[htpb]
  \centering
  \includegraphics[width=0.8\linewidth]{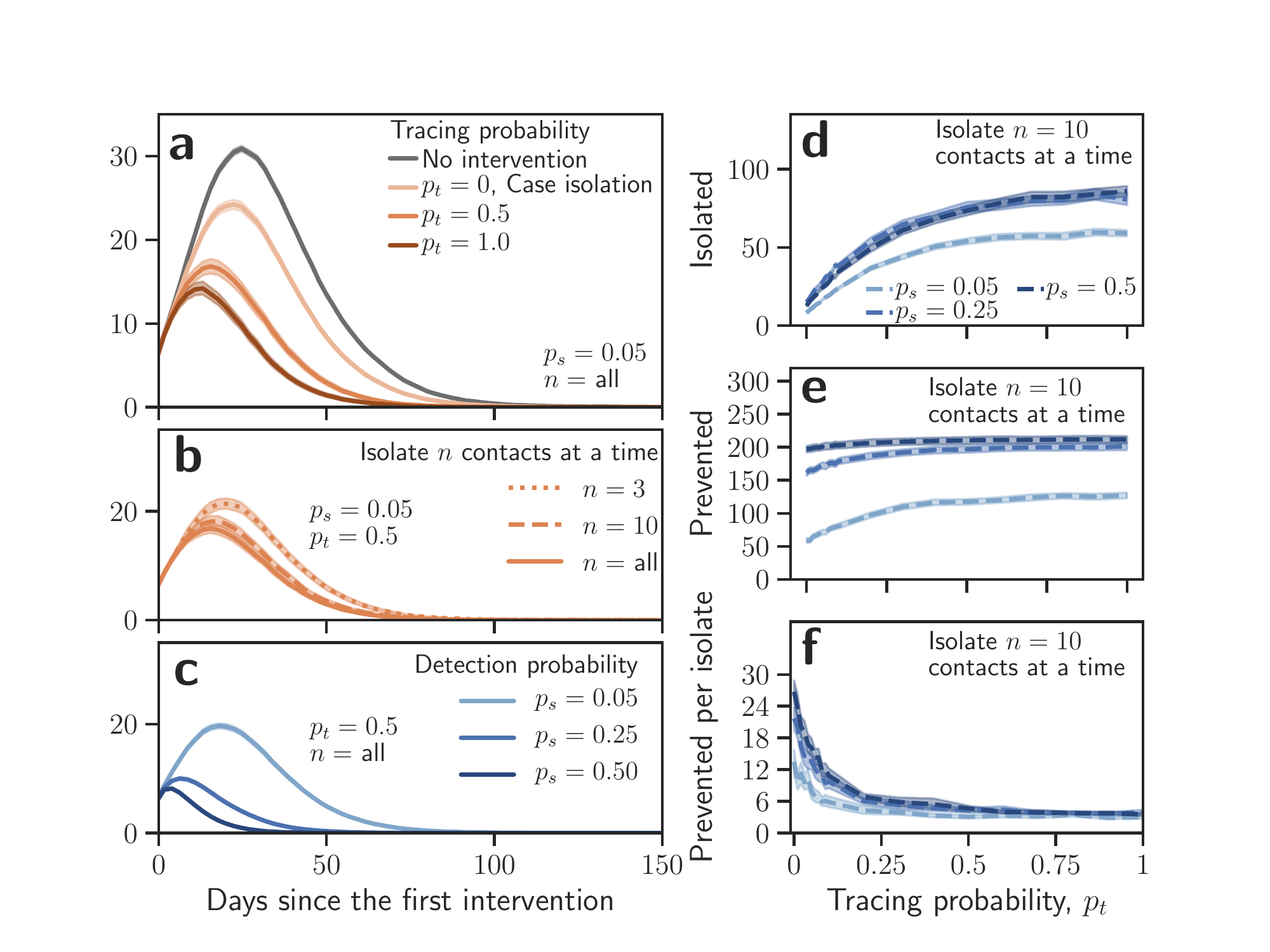}
  \caption{
    Effectiveness of contact tracing for the student physical contact network. 
    An infected node is discovered and isolated with probability $p_s$. 
    Contact tracing isolates the most frequent $n$ close contacts in the contact list.
    We isolate $n=3, 10$, or all close contacts, which are indicated by ``$n=3$'', ``$n=10$'', or ``all'', respectively.
    {\bf a}. The contact tracing reduces the peak of infections more than case isolation.
    {\bf b}. Even if we do not trace and isolate all but few nodes, it is as effective as isolating all contacts.
    {\bf c}. The effectiveness is more pronounced when we can detect more infected nodes.
    {\bf d}--{\bf f}. Contact tracing isolates more nodes and prevents more cases as we trace more contacts.
    Contact tracing is not efficient when tracing probability is large. Although contact tracing is highly effective and efficient, massive contact tracing may have a diminishing return.
    Each point indicates the average value for $1,000$ simulations.
    The translucent band indicates the 95\% confidence interval estimated by a bootstrapping with $10^4$ resamples. 
  }%
  \label{fig:sensible_dtu_effectiveness}
\end{figure}

Our simulation shows that case isolation alone reduces the peak of infections by roughly 15\% (\figref{sensible_dtu_effectiveness}a). 
Contact tracing lowers the peak by about 50\% even though the network does not exhibit strong heterogeneity (\figref{sensible_dtu_effectiveness}a).
Moreover, tracing and isolating few traced contacts has comparable effectiveness to isolating all close contacts (\figref{sensible_dtu_effectiveness}b).
The peak can be further reduced by contact tracing when we can detect more infected nodes, \ie increasing testing capacity (\figref{sensible_dtu_effectiveness}c).
Contact tracing has a marked diminishing return; as tracing probability $p_t$ increases, contact tracing isolates more nodes but prevents nearly the same number of cases (\figref{sensible_dtu_effectiveness}d--f).
Still, contact tracing yields a high benefit; it prevents at least roughly five cases per isolation.
In sum, our results suggest that even when the network is homogeneous and densely connected, a small amount of contact tracing may be able to curve the spreading efficiently.


\subsection{Analytical analysis of contact tracing on networks with arbitrary degree sequence}

\begin{figure*}
\centering
\includegraphics[width=0.45\linewidth]{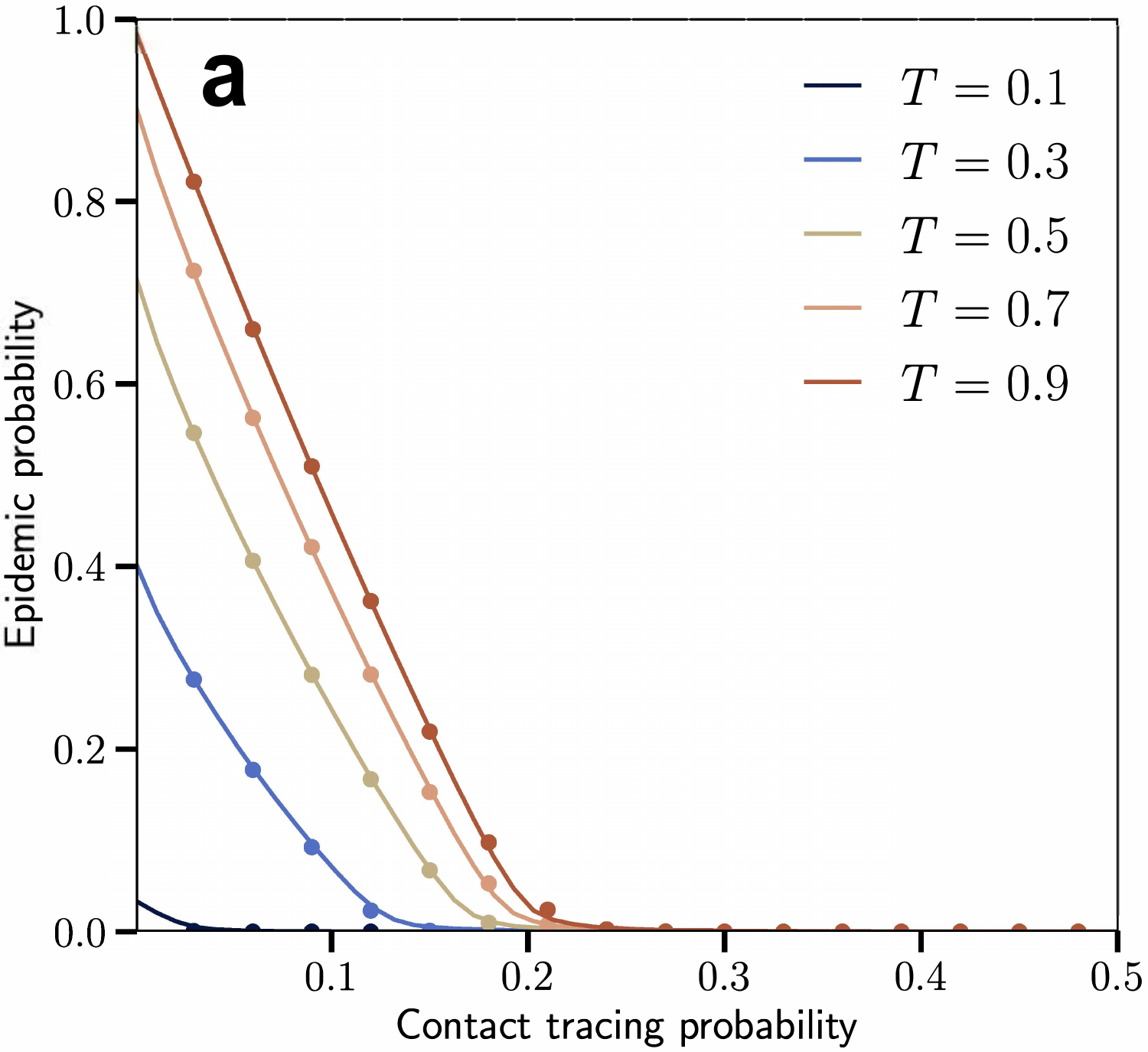}
\includegraphics[width=0.45\linewidth]{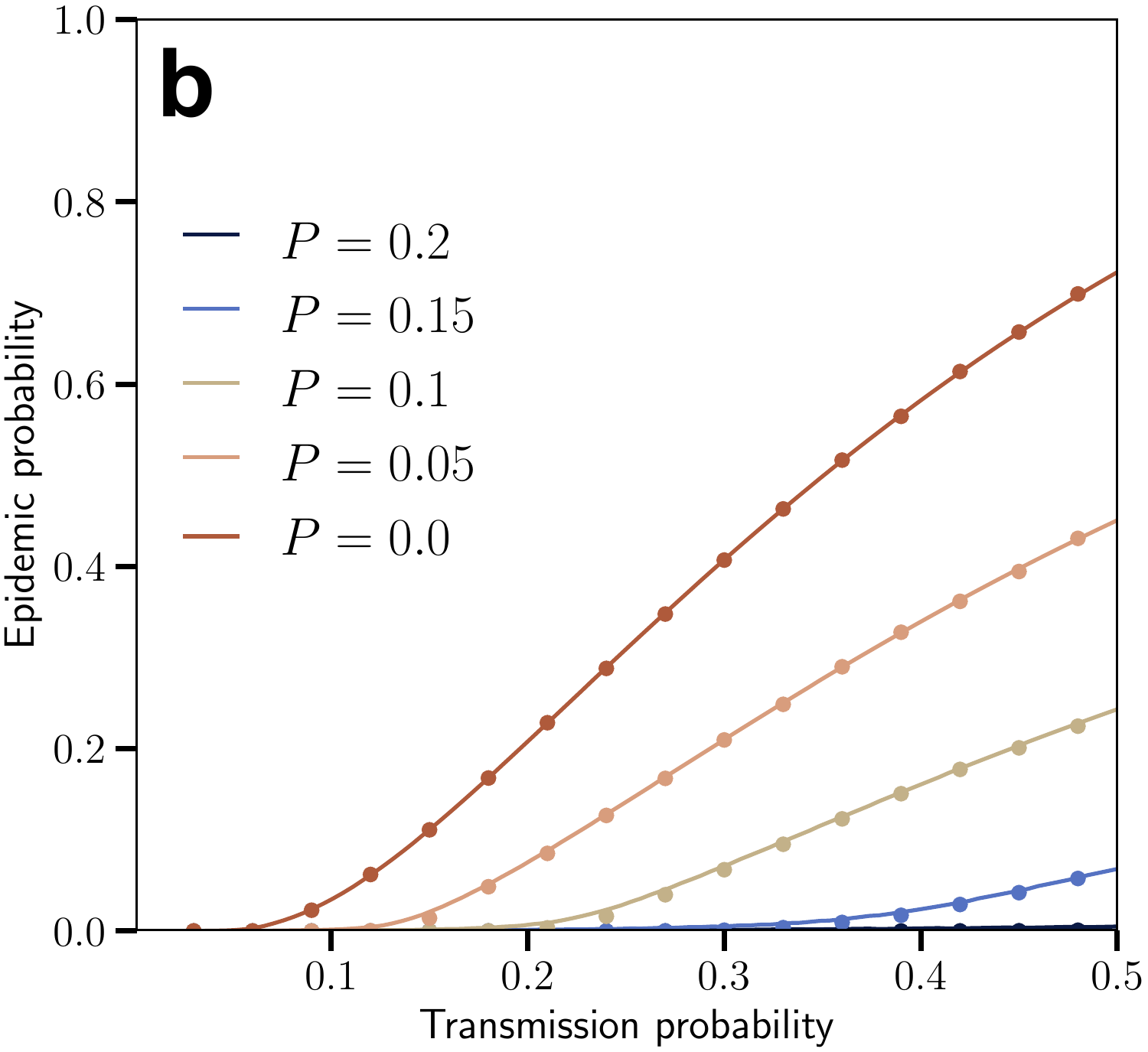}
\caption{Control of an outbreak using contact tracing in heterogeneous networks. We use the BA network, where each of 250,000 nodes has a degree at least 2 and attempt to control the spread of a disease with transmissibility $T$ using tracing probability $P$. Markers show the average of 100 Monte Carlo simulations, and solid lines show the results of our analytical formalism. {\bf a}. The probability of sustained transmissions goes down monotonically with more contact tracing, but without undergoing the usual sharp epidemic transition. Unlike in mass-action models, there are diminishing returns to contact tracing: While it efficiently identified super-spreading events in heterogeneous networks, the epidemic eventually localizes around low degree nodes which are harder to protect with contact tracing alone. {\bf b}. The regime of smeared epidemic transition increases with the frequency of contact tracing. At a high frequency of contact tracing, we find the probability of sustained transmission remains low even for high values of transmissibility well-beyond the epidemic threshold.}
\label{fig:epidemic_probability}
\end{figure*}

Let us investigate how much contact tracing would be necessary to prevent an outbreak.
We calculate the epidemic probability---the probability of sustained transmission of disease---for networks with an arbitrary degree distribution under contact tracing based on a branching process formalism (see Epidemic probability in Methods for the derivation of the probability).
We consider a contact network of people, where a disease is transmitted from an infected person $i$ (\ie parent) to susceptible person $j$ (\ie offspring) with transmission rate $T$.
The parent is identified and isolated with probability $P=p_s p_t$ (\ie tracing probability) from its offspring $j$ by contact tracing.

Our analytical solution (see Epidemic probability in Methods), as well as a numerical simulation (\figref{epidemic_probability}), demonstrates that increasing the tracing probability $P$ can control an epidemic and stop any possibility of sustained transmission while showing a diminishing return of contact tracing. Notably, we find a smooth epidemic threshold in $P$, which is distinct from the usual sharp epidemic threshold observed over $T$. This phenomenology can be understood by considering \textit{who} gets targeted by contact tracing. 
Effective execution of contact tracing detects transmissions events from an individual with a probability proportional to $k^2$, where $k$ is the degree of the individual.
Consequently, as we increase the frequency of contact tracing, we not only reduce the number of transmissions but do so by only allowing transmissions to occur around relatively small degrees. Therein lies the power of contact tracing on heterogeneous networks, it reduces the size of the epidemic and localizes it around nodes of lower degrees; reducing both the total number of infections and the frequency of super-spreading events.

\section{Discussion}\label{sec:discussion} 

We show that contact tracing leverages two sampling biases arising from the heterogeneity in the number of individual's contacts.
Our theoretical and simulation analyses indicate that contact tracing can be a highly effective and efficient strategy even when it is not performed on a massive scale, as long as it is strategically performed to leverage the sampling biases.
Furthermore, contact tracing can be more cost-efficient than case isolation in terms of the number of prevented cases per isolation, in particular when detecting infection is difficult.
The effectiveness and efficiency hinge upon the fact that backward tracing can detect super-spreading events exceptionally well.
Therefore, we argue that (i) even when massive contact tracing is not feasible, it may still be worth to implement contact tracing, (ii) not all contact tracing protocols are equal---it is crucial to implement the protocols that leverage the presented biases, and (iii) the ``cheaper'' contact tracing offered by digital contact tracing may hold even greater potential than previously suggested~\cite{hellewell2020feasibility}.

In the context of digital contact tracing, our results show the need for (i) backward contact tracing that aims to identify the parent of a detected case and (ii) deep contact tracing to notify other recent contacts of the traced nodes. 
Current implementations of digital contact tracing, including the Apple and Google partnership~\cite{apple-google} 
and the DP-3T proposal~\cite{dp3t}, notify the contacts 
of an infected individual about the risk of infection. 
However, they neglect that one of these previous contacts is likely the source of infection (\ie parent) who might be infecting others. 
We show that multiple notifications are particularly indicative of the parent and can be potentially leveraged for better intervention strategies. 
Therefore, we urge the consideration of a multi-step notification feature that can fully leverage the sampling biases arising from the heterogeneity in the contact network structure. 

An implementation of our model does not necessarily require any compromise in terms of privacy or decentralization of the contact tracing protocol itself~\cite{cho2020contact}. 
One could also imagine a hybrid approach, where, deep contact tracing is undertaken using a centralized database when a given device has been notified more than a certain amount of time.
The benefits of such network-based contact-tracing could be significant, especially if accompanied by serious educational efforts for users to explain the rationale behind the intervention and the importance of their own role in our social network.

There are several caveats to be considered.
First, diagnostic tests and isolation are assumed to be instantaneous in our simulations.
A huge delay may degenerate the effectiveness of preventive measures, in particular case isolation in which immediate isolation is crucial. 
Second, we assume that every individual has an equal probability of infection and isolation, which, however, may vary depending on demographics. The heterogeneity in the probabilities may hinder the effectiveness of opt-in contact tracing strategies. For example, it is possible that a virus is constantly sourced from people who refuse contact tracing~\cite{sarah2020contact,steven2020time} or who traveled from a different country that does not share contact data.

Even with the aforementioned limitations, our results suggest that contact tracing has a larger potential than commonly considered.
Because the effectiveness hinges upon the ability to reach the ``source'' of infection, our results underline the importance of strategic contact tracing protocols.



\section{Methods}\label{sec:methods}

\subsection{Data}

We use the dataset collected in the Copenhagen Network Study \cite{sapiezynsky2019interaction} to construct the temporal network of physical contacts between students in a university. 
The data set contains information on the physical contacts between more than 700 students in a university estimated by Bluetooth signal strength. 
We remove all individuals from the data that have a valid Bluetooth scan in less than 60\% of the observation period.
Then, we regard that two individuals $i$ and $j$ had a contact if $i$ or $j$ received the Bluetooth scans from the other with the signal strength more than $-75$dB. We note that one receives the signal strength at approximately 1m distance from the device \cite{sekara2014thestrength}.
These steps resulted in a cohort of $N=567$ individuals with contact data for $28$ days with 5 min resolution.

\subsection{Simulating epidemic spreading}

We simulate the SIR model for the static contact networks and people-gathering networks using the \texttt{EoN} package~\cite{kiss2017mathematics}, with transmission rate $T = 0.25$, recovery rate $\gamma = 0.25$, and initial seed fraction $\rho = 10^{-3}$. 

For the student contact network, we simulate the SEIR model with the parameters used in studies on the COVID-19 disease~\cite{zhang2020science}: expected infectious and incubation periods are set to 5 and 1 days, respectively.
The transmission rate of the COVID-19 highly varies across case studies and estimation methods~\cite{hebert2020beyond,hellewell2020feasibility}. 
One expects that, in any closed population with dense contacts, between 20\% to 60\% of the population are infected~\cite{hebert2020beyond}. 
Therefore, we use a transmission rate $0.5$ day$^{-1}$ to produce outbreaks that reach 50\% of the population, which is close to the worst-case scenarios that might be expected on a university campus
We randomly choose 1\% of the total population as initially infected nodes at time $t_0$, where $t_0$ is chosen randomly in the first 28 days.
The epidemic spreading process may take longer than the days recorded in the contact data (\ie 28 days).
Therefore, following a previous study~\cite{valdano2015}, we assume that the contacts on the first day ensue after the last day. 



\subsection{People--gathering networks}
In the people-gathering network, a person-node is connected to a gathering-node if he/she joined the gathering.
The degree of a person implies how mobile the person is across diverse sets of gatherings, and the degree of a gathering indicates the number of participants for the gathering. 
Denoted by $G_0(x)$ and $F_0(x)$ the generating functions for the degree distributions of persons and gatherings, respectively, which are defined as
\begin{eqnarray}
	G_0(x) &=& \sum_k p_k x^k,\\
	F_0(x) &=& \sum_k q_k x^k.
\end{eqnarray}
The transmission event happens from a person to others \textit{via} a gathering. 
When we trace a gathering from a person, a gathering with $k$ participants is $k$ times more likely to be sampled than the gathering with only one person. Therefore, the excess size of the gathering is generated by 
\begin{equation}
	F_1(x) = \frac{F'_0(x)}{F'_0(1)} = \frac{1}{\sum_k k q_k} \sum_k k q_k x^{k-1}.
\end{equation}
The probability distribution of the number of one's neighbors through gatherings is given by $G_0(F_1(x))$.
Because larger gatherings would produce more infections and thus more likely to be traced, the number of participants of the gathering except for the original spreader and the isolated individual is given by the probability generating function
\begin{equation}
	F_2(x) = \frac{F'_1(x)}{F'_1(1)} = \frac{1}{\sum_k (k^2 -k) q_k} \sum_k k (k-1) q_k x^{k-2}. 
\end{equation}
In other words, contact tracing samples a gathering with $k$ participants with probability roughly proportional to $k^2$.
Therefore, as is the case for people contact networks, contact tracing is effective at identifying super-spreading events and prevent numerous further disease transmission events.

\section{Epidemic probability}

We calculate the probability that the contact tracing stops the spreading of disease.
To keep the analysis simple, we assume that every newly infected node has a probability $P$ to lead to its parent node and we can prevent the infections to all of the parent's grandchildren by notifying the infected node. 

The probability of epidemics is determined by the offspring distributions, \ie number of nodes to which an infected node spreads the disease.
We note that the offspring distribution depends on how we sample nodes due to the sampling biases (see Results).
Specifically, if we sample infected nodes at random or by following a random transmission, the offspring distributions are given by generating functions
\begin{align}
R_0(x) = G_0(Tx+(1-T)) = \sum _k r_k x^k \quad \textrm{or} \quad  
R_1(x) = G_1(Tx+(1-T)) = \sum _k q_k x^k,  
\end{align} 
respectively, where $T$ is the probability of transmitting disease through an edge, and $r_k$ and $q_k$ are the probabilities of having $k$ offsprings, respectively.

With contact tracing, the offsprings of a parent can continue the spreading process if and only if successful contact tracing does not take place for all the offsprings, which occurs with probability $(1-P)^k$.
Therefore, the nodes sampled by following a random transmission have the offspring distribution given by
\begin{equation}
\overline{R} _1(x,y) = \sum_k q_k\bigg\lbrace(1-P)^kx^k + \left[1-(1-P)^ky^k\right] \bigg\rbrace,
\end{equation}
where the $\overline R_1$ denotes the $R_1$ under contact tracing, and $\overline R_0$ is the analogous function for $R_0$.
We have distinguished standard transmissions (counted with the variable $x$) from transmissions that occurred but are isolated quickly enough by contact tracing to stop the transmission tree (counted with the variable $y$). 
This gives us a way to calculate the coefficients $r_k$ of $\overline{R}_0(x,1)$ which specify the distribution of successful branching events in the transmission tree (\ie those that can continue spreading).

The probability $u$ that transmission to a node without contact tracing around the parent does \textit{not} lead to sustained transmission is given by the self-consistency condition
\begin{equation}
u = \overline{R}_1(u,1) \; ,
\label{eq:u}
\end{equation}
where the right-hand side gives the probability that the offsprings also do not lead to sustained transmission (1 if contact tracing occurs, and $u$ otherwise). The probability of an epidemic is then the probability that at least one transmission around patient leads to sustained transmission, or
\begin{equation}
\Pi = 1-\overline{R}_0(u,1) \; .
\end{equation}

\bibliography{main.bib}

\noindent\textbf{Acknowledgements}\\
The authors would like to thank M.~Girvan, J.~Lovato, and other organizers of the Net-COVID program, which initiated the project. We also thank A.~Allard, C.~Moore, E.~Moro, A.~S.~Pentland, and S.~V.~Scarpino for helpful discussions.
L.~H.-D. acknowledges support from the National Institutes of Health 1P20 GM125498-01 Centers of Biomedical Research Excellence Award. S.~K. and Y.-Y.~A. acknowledges support from the Air Force Office of Scientific Research under award number FA9550-19-1-0391.\\

\noindent\textbf{Author contributions}\\
Y.-Y.~A. conceived the research. E.~M., S.~K., L.~H.-D. and Y.-Y.~A. performed the numerical simulations. L.~H.-D. and Y.-Y.~A. conducted the mathematical analysis. All authors participated in the analysis and interpretation of the results as well as the writing of the manuscript. 

\noindent\textbf{Competing interests}\\
We have no competing interests.


\end{document}